\documentclass{aastex631}

\usepackage{subfigure}
\usepackage{CJK}

\submitjournal{ApJS}
\shorttitle{AASTeX v6.3.1 Sample article}
\shortauthors{YongLin YU et al.}
\graphicspath{{./}{figures/}}

\begin{document}
\begin{CJK*}{UTF8}{gbsn}
\title{Two-element interferometer for millimeter-wave solar flare observations}

\correspondingauthor{FaBao Yan ({\CJKfamily{gbsn}严发宝})}
\email{yanfabao2022@163.com}

\author[0000-0003-1208-6764]{YongLin Yu ({\CJKfamily{gbsn}于永林}) }
\affiliation{Laboratory for Electromagnetic Detection (LEAD),Institute of Space Sciences, Shandong University \\ Weihai 264209, People's Republic of China;yanfabao2022@163.com}

\author{Shuo Xu ({\CJKfamily{gbsn}许硕})}
\affiliation{Laboratory for Electromagnetic Detection (LEAD),Institute of Space Sciences, Shandong University \\ Weihai 264209, People's Republic of China;yanfabao2022@163.com}
\affiliation{School of Mechanical, Electrical \& Information Engineering, Shandong University \\ Weihai 264209, People's Republic of China}

\author{Lei Zhang ({\CJKfamily{gbsn}张磊})}
\affiliation{Laboratory for Electromagnetic Detection (LEAD),Institute of Space Sciences, Shandong University \\ Weihai 264209, People's Republic of China;yanfabao2022@163.com}

\author[0000-0002-8826-8006]{ZiQian Shang ({\CJKfamily{gbsn}尚自乾})}
\affiliation{Laboratory for Electromagnetic Detection (LEAD),Institute of Space Sciences, Shandong University \\ Weihai 264209, People's Republic of China;yanfabao2022@163.com}

\author{ChengLong Qiao ({\CJKfamily{gbsn}乔成龙})}
\affiliation{Laboratory for Electromagnetic Detection (LEAD),Institute of Space Sciences, Shandong University \\ Weihai 264209, People's Republic of China;yanfabao2022@163.com}
\affiliation{School of Mechanical, Electrical \& Information Engineering, Shandong University \\ Weihai 264209, People's Republic of China}

\author{ShuQi Li ({\CJKfamily{gbsn}李舒琪})}
\affiliation{Laboratory for Electromagnetic Detection (LEAD),Institute of Space Sciences, Shandong University \\ Weihai 264209, People's Republic of China;yanfabao2022@163.com}
\affiliation{School of Mechanical, Electrical \& Information Engineering, Shandong University \\ Weihai 264209, People's Republic of China}

\author{Zhao Wu ({\CJKfamily{gbsn}武昭})}
\affiliation{Laboratory for Electromagnetic Detection (LEAD),Institute of Space Sciences, Shandong University \\ Weihai 264209, People's Republic of China;yanfabao2022@163.com}

\author{YanRui Su ({\CJKfamily{gbsn}苏艳蕊})}
\affiliation{Laboratory for Electromagnetic Detection (LEAD),Institute of Space Sciences, Shandong University \\ Weihai 264209, People's Republic of China;yanfabao2022@163.com}
\affiliation{School of Mechanical, Electrical \& Information Engineering, Shandong University \\ Weihai 264209, People's Republic of China}

\author{HongQiang Song ({\CJKfamily{gbsn}宋红强})}
\affiliation{Laboratory for Electromagnetic Detection (LEAD),Institute of Space Sciences, Shandong University \\ Weihai 264209, People's Republic of China;yanfabao2022@163.com}

\author[0000-0001-6449-8838]{Yao Chen ({\CJKfamily{gbsn}陈耀})}
\affiliation{Laboratory for Electromagnetic Detection (LEAD),Institute of Space Sciences, Shandong University \\ Weihai 264209, People's Republic of China;yanfabao2022@163.com}

\author[0000-0002-4451-7293]{FaBao Yan ({\CJKfamily{gbsn}严发宝})}
\affiliation{Laboratory for Electromagnetic Detection (LEAD),Institute of Space Sciences, Shandong University \\ Weihai 264209, People's Republic of China;yanfabao2022@163.com}
\affiliation{School of Mechanical, Electrical \& Information Engineering, Shandong University \\ Weihai 264209, People's Republic of China}



\begin{abstract}

In this paper, we present the design and implementation of a two-element interferometer working in the millimeter wave band (39.5 GHz$\sim$40 GHz) for observing solar radio emissions through nulling interference. The system is composed of two 50-cm aperture Cassegrain antennas mounted on a common equatorial mount, with a separation of 230 wavelengths. The cross-correlation of the received signals effectively cancels the quiet solar component of the large flux density ($\sim$3000 sfu) that reduces the detection limit due to atmospheric fluctuations. The system performance is obtained as follows: the noise factor of the AFE in the observation band is less than 2.1 dB, system sensitivity is approximately 12.4 K ($\sim$34sfu) with an integration time constant of 0.1 ms (default), the frequency resolution is 153 kHz, and the dynamic range is $\ge$30 dB. Through actual testing, the nulling interferometer observes a quiet sun with a low level of output fluctuations (of up to 50 sfu) and has a significantly lower radiation flux variability (of up to 190 sfu) than an equivalent single-antenna system, even under thick cloud cover. As a result, this new design can effectively improve observation sensitivity by reducing the impact of atmospheric and system fluctuations during observation.

\end{abstract}


\section{Introduction} \label{sec:Introduction}

Solar radio emission can be significantly affected by the atmosphere of the Earth when transmitting to the Earth’s surface, such as the turbulence of the atmosphere and absorption of vapor and oxygen molecules ( \cite{2021SoPh..296....9M};\cite{2014ApJ...790...67L};\cite{2016ApJ...831..154M} ). Therefore, the antennas receive signals with a "dither" component, which corresponds to a noise signal that varies randomly in the radiation intensity. Thus, the sensitivity of an observing system at millimeter wavelengths deteriorates severely due to the weather (\cite{2020JSWSC..10....7C};\cite{1957IRE..5....12K};\cite{2020ppot.book.....P};\cite{2020FrASS...7...79C}). which makes it difficult to detect weaker radio bursts.

Observation of small solar flares can be achieved through the use of large antennas with a narrow beamwidth at half power. This approach enables the focusing of electromagnetic radiation to a precise and concentrated region, thereby increasing the sensitivity of the detector. Additionally, in the dual-antenna correlation technique, also known as a nulling interferometer, the visibility of the interferometer output during quiet sun observations is heavily reliant on the spacing between the antennas. At specific antenna spacings, the visibility may diminish to zero, so the quiet sun emission can be canceled out (\cite{1985PASJ...37..163N}).When the observation is performed with a large-aperture antenna, the size of the spatial resolution is proportional to the observation wavelength and inversely proportional to the antenna diameter. Its beam width is narrower than the solar disk (approximately $0.5^{\circ}$), the received flux density decreases, the signal-to-noise ratio increases, and therefore, the absolute value of atmospheric fluctuations also decreases in the same proportion(\cite{2019AstBu..74..221K}). For example, Shimabukuro used an antenna with a half-power beam width of 3 arcmins to observe solar flares at 90 GHz in 1970 (\cite{1970SoPh...15..424S}). The 14-meter radio telescope (RT-14) at the MRO Observatory of Aalto University, Finland, has a beam width of 2.4 arcmins when observing an 8 mm signal(\cite{2018SoPh..293..156K}).The Cagliari Astronomical Observatory used the Medicina 32 m and SRT 64 m antennas for solar imaging observations and obtained approximately 170 maps of the entire solar disk in the 18GHz$\sim$26 GHz band, with beam widths of 2.1 arcmins and 1.5 arcmins for 18 GHz and 26 GHz solar observations, respectively, when using the Medicina 32 m antenna. The beam widths of the 18 GHz, 24 GHz and 25 GHz solar observations with the SRT 64 m antenna are 1.02 arcmin, 0.78 arcmins, and 0.75 arcmins, respectively(\cite{2022SoPh..297...86P};\cite{2017A&A...608A..40P}).When correlated by two antennas (nulling interferometer) to compensate for the quiet solar background radiation, the solar flare can be considered a point source signal compared to the solar disk, and the nulling interferometer can largely reduce the fluctuations in the overall quiet solar flux density(\cite{2005ITMTT..53.1168L}).The sensitivity of the nulling interferometer is higher than that of the single-antenna radiometer with the same antenna diameter. The nulling interferometer detects the burst of small flares on the sun with two small antennas with beam widths covering the whole sun; however, if each antenna is installed separately, the effective baseline distance varies with the sun's time angle(\cite{2016AmJPh..84..249K}).Therefore, the two antennas must be mounted in the same plane to ensure a stable effective baseline. A few applications of the nulling interferometer specifically for solar radio observations have been reported, such as the 80 GHz radiometer built at the Nobeyama Observatory in 1984, which was the first use of the nulling interferometer for solar outburst monitoring, using two 25 cm diameter antennas mounted on the same mount. Many flare-outbursts were monitored during the shutdown until 2018(\cite{1984TokyoAOr..20..327S}). A nulling interferometer for observing solar flares at 90 GHz was built at the University of Bern, Switzerland, in 1999(\cite{1999Bern..L}).The performance details and comparisons of the above two sets of nulling interferometers and the presented system are shown in Table 1.

According to the gyrosynchrotron mechanism, the spectral properties of radio bursts in the centimeter-millimeter band can be used to diagnose the magnetic field information of the flaring region in the lower corona(\cite{2011SoPh..268..165K}). However, currently, there are only observations at only several discrete frequencies beyond $\sim$20 GHz. A 35 GHz$\sim$40 GHz solar radio dynamic spectrum observation system exists at the Chashan Solar Radio Observatory (CSO) of Shandong University(\cite{2022ApJS..258...25S}), but the system receives signals by a single antenna and therefore is affected by atmospheric dither and weather. To decrease the solar microwave signal noise caused by atmospheric dither, improve the anti-turbulence capability and sensitivity of flare signal measurements, and detect small solar flares in the 40 GHz band, a two-element interferometer for millimeter wave solar radio bursts has been developed, which was installed at the Chashan Solar Radio Observatory (CSO) and tested for solar radio observation. We use two 50 cm aperture, circularly polarized Cassegrain-type antennas mounted on the same planar mount at a baseline distance of 230 wavelengths to receive the solar radiation signals and conduct cross-correlation processing to cancel the radiation of the quiet sun, reduce the fluctuation of the entire solar radiation, and monitor the outbreak of small solar flares in the 40 GHz band with high sensitivity. Section 2 introduces the system-related principles, Section 3 describes the design of the whole system, Section 4 shows the performance of the system, Section 5 describes the approximate calibration of the system, and finally, the conclusion and discussion are given.

\begin{deluxetable*}{lccccBcccccBcccc}
	\tablenum{1}
	\tablecaption{The performance of the proposed system and other nulling interferometers for flare observation\label{tab:messier}}
	\tablewidth{0pt}
	\centering
	\tablehead{ \colhead{Parameter} & \colhead{90 GHz(NIOS)}  & \colhead{80 GHz(NoRP)} & \colhead{40 GHz}}
\startdata
	Observation Wavelength($\lambda$) & 3.3$\thinspace mm(90\thinspace GHz)$  & $3.75\thinspace mm(80\thinspace GHz)$ & $7.5\thinspace mm(40\thinspace GHz)$\\
	Polarization       & $Linear/Horizontal$ & $Circular Polarization$ & $Circular Polarization$ \\
	System Temperature & $3300\thinspace (2300) \thinspace K$ & $\sim 3000 \thinspace K$ &  $\sim 1247 \thinspace K$ \\
	Integration Time Constant & $31\thinspace ms$ & $0.2\thinspace s$ & $0.1\thinspace ms$\\
	Bandwidth & $500\thinspace MHz(single sideband)$ & $400\thinspace MHz(double sideband)$ & $500\thinspace MHz(single sideband)$ \\
	Radiometric Sensitivity & $0.6\thinspace (0.4)\thinspace K(20-35\thinspace sfu)$ & \ $\sim 15\thinspace sfu$ & \ $\sim 12.4\thinspace K\thinspace (34\thinspace sfu) $ \\
	Antenna Aperture & $-$ & $25\thinspace cm$ & $50\thinspace cm$ \\
	Antenna HPBW & $0.9\thinspace deg$ & $1.08\thinspace deg(65arcmin)$ & $1.2\thinspace deg$ \\
	Antenna Spacing &  $\sim 1.15\thinspace m(350 \lambda)$ &  $\sim 1.24\thinspace m(330 \lambda)$ &  $\sim 1.73\thinspace m(230 \lambda) $ \\	
\enddata
\end{deluxetable*}

\section{The related principle of the system} \label{sec:style}

The system consists of two small antennas with beamwidths covering the entire solar disk for interference. Two small antennas are mounted on a common mount to keep the effective baseline $D$ of the antenna from changing with the position of the sun (Figure 1). The RF signals output from the two antennas are downconverted (sharing the same local oscillator source to maintain phase consistency) and then fed into a digital correlator for correlation. The phase difference $\varphi$ of the signals from the two antennas is (\cite{2007MeScT..18...41G})

\begin{equation}
	\varphi =2\pi \mu \theta =\frac{2\pi D\sin \theta }{\lambda },
\end{equation}

where $\lambda$ is the observed wavelength and $\frac{\pi}{2}-\theta$ is the angle between the baseline and the solar direction. The amplitude and phase of the observed sun are obtained from the correlator (\cite{2000PASJ...52..393O}) as

 $F=\sqrt{R_{C}^{2}+  R_{S}^{2}  } \quad and \quad \varphi =\tan^{-1} \left ( \frac{R_{S} }{R_{C} }  \right )$
, respectively, where $R_{C} \sim A\left ( \theta  \right ) \cos \left ( \varphi  \right )$   and $R_{S} \sim A\left ( \theta  \right ) \sin  \left ( \varphi  \right )$ .

\begin{figure}[htbp]
	\centering
	\begin{minipage}[t]{0.48\textwidth}
		\centering
		\includegraphics[width=7cm]{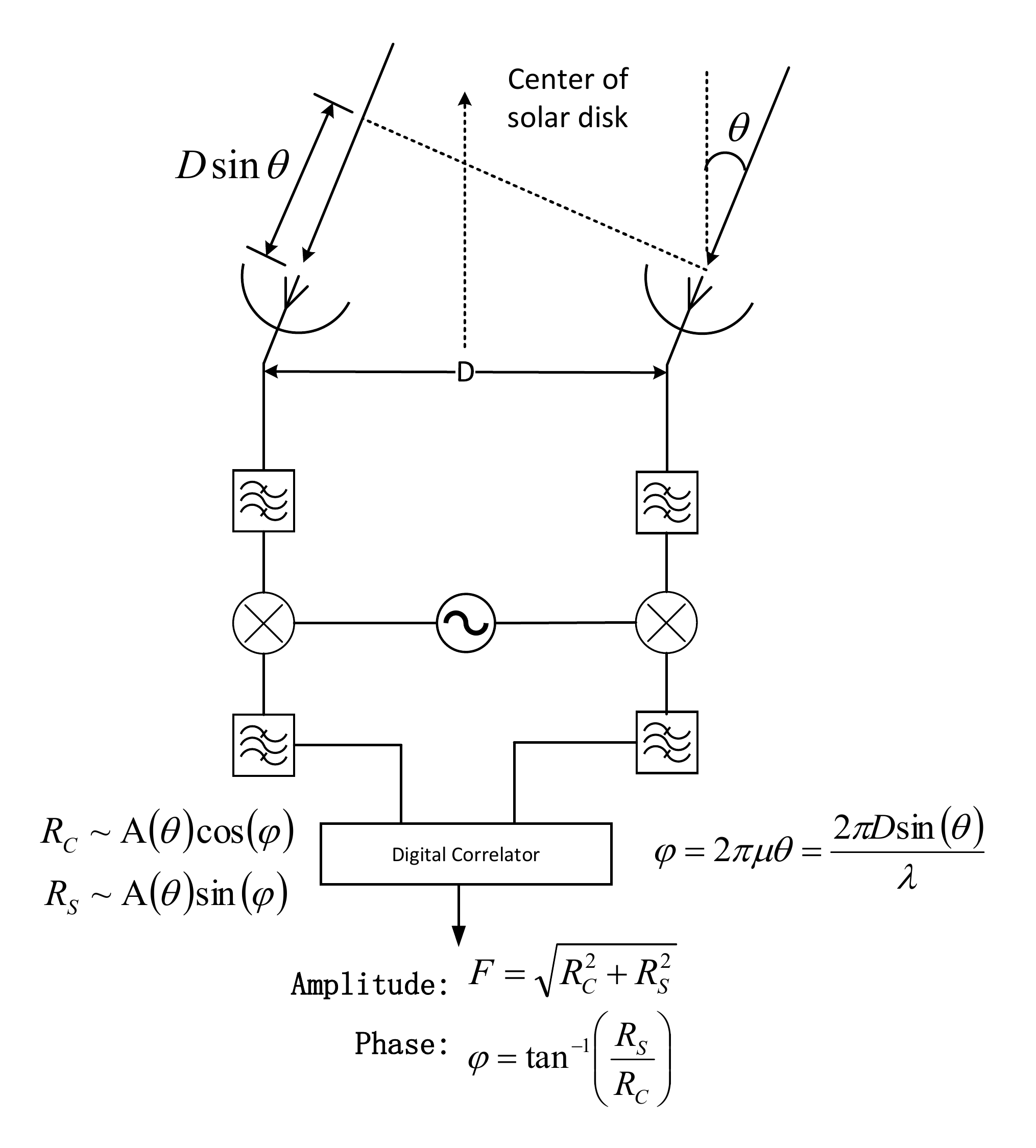}
		\\ \caption{Simplified block diagram of the two-element interference system. The solar radiation signal is received by two small antennas (H and E) mounted on a coplanar mount. The RF signal is filtered and downconverted (sharing the same local oscillator source) and then output to the digital correlator \label{fig:general}}
	\end{minipage}
	\begin{minipage}[t]{0.48\textwidth}
		\centering
		\vspace{-22.5em}
		\setlength{\abovecaptionskip}{2.3em}
		\includegraphics[width=8.5cm]{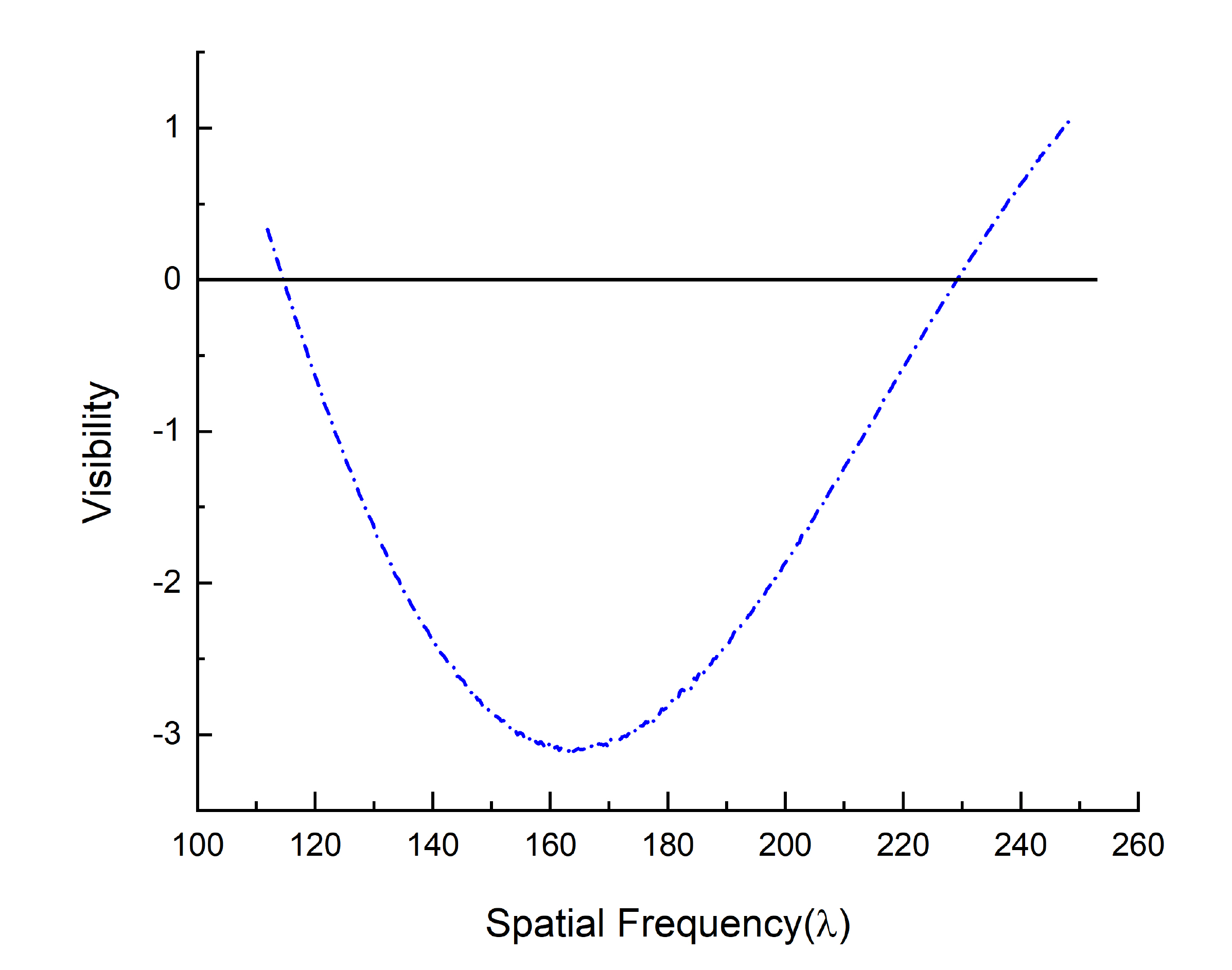}
		\\ \caption{Curve of the visibility function as a function of the spatial frequency output from the two-element interferometer. When conducting solar observations, the baseline length corresponding to the second zeroing point is chosen, corresponding to an observation baseline of approximately $230\lambda$ (the slide module can be microadjusted)  \label{fig:general}}
	\end{minipage}
\end{figure}

Assume a one-dimensional source with $ \ell = \sin \theta  $ and projection baseline $\mu$. For a distant source with a small angular range, such as ground-based observation of the Sun, a small-angle approximation can be used with $\ell = \sin \theta\approx \theta $, the spatial frequency $\mu = \frac{b\cos \theta _{0}}{\lambda }  $, and the position-referenced phase $\theta _{0} $ (\cite{2004ITGRS..42.1677C}). From Fourier theory, it is known that the symmetric function consists of only cosine Fourier series, so the visibility function is (\cite{1998ITGRS..36..680C}).

\begin{equation}
	\mathrm {V} \left ( \mu  \right )   =\int  _{\frac{-\Delta \theta }{2} }^{\frac{\Delta \theta }{2} } \mathrm {A}\left ( \theta  \right )\mathrm {I}\left ( \theta  \right )\cos \left ( 2\pi \mu \theta  \right ) d\theta  ,
\end{equation}

where $\mathrm {A}\left ( \theta  \right )$ is the antenna orientation map and $\mathrm {I}\left ( \theta  \right )$ is the intensity profile of the one-dimensional source. As the antenna direction map stereo angle is much wider than the target source, the observed flux density $\mathrm {B}  \left ( \theta ,\phi  \right )  \Omega_{s}  $ is reduced to $\mathrm {I} \left ( \theta  \right ) \bigtriangleup \theta  $, where $\mathrm {I}=- \mathrm {B} $, and the above equation can be reduced to

\begin{equation}
	\mathrm {V}\left ( \mu  \right ) = \left | \mathrm {V}    \right |\frac{\sin \pi \mu \bigtriangleup \theta }{\pi \mu \bigtriangleup \theta}    ,
\end{equation}

From the above equation, the fringe function amplitude tends to 0 when $\pi \mu \Delta \theta \to \pi  $ and the solar view angle $\Delta \theta $ is approximately $0.5^{\circ}$ (aphelion: $0.532^{\circ}$, perihelion: $0.542^{\circ}$). Therefore, the minimum baseline required to observe the Sun is $b=\frac{\lambda }{\Delta \theta } $, and the minimum baseline corresponding to the observation band 39.5 GHz $\sim$ 40 GHz ($\lambda \in \left ( 7.5mm,7.595mm \right ) $) is 859 mm $\sim$ 872 mm. When $\pi \mu \Delta \theta \to 2\pi  $, the minimum baseline corresponding to the observation band is 1718 mm $\sim$ 1740 mm.

The two-element interferometer is used to track the sun in a variable baseline (850 mm$\sim$1890 mm), and the visibility function of the two-element interferometer output as a function of spatial frequency can be obtained as shown in Fig. 2. In the baseline (850 mm$\sim$1890 mm), the corresponding spatial frequencies are $110\lambda \sim 250\lambda $. The baseline length corresponding to the second zeroing point is chosen when the two-element interferometer is used for solar observation. The observation baseline is $230\lambda$ (with some fluctuations depending on the observation frequency).

When the baseline of the observed quiet Sun is $230\lambda$, the value of the visibility function is zero, so the radiation from the quiet Sun can be compensated, and the fluctuations caused by the Earth's atmosphere to the large flux density are largely reduced. Additionally, since flares occur in a small region on the solar disk, the flare-related output is not affected, and the system has a high sensitivity to solar flares.

When a part of the sun's surface is obscured by clouds, the solar radiation signal received by the ground changes, causing fluctuations, and when two-element interference is used, the visibility function changes as follows.

\begin{equation}
	\Delta \mathrm {F}\left ( \mu  \right ) =- \Delta \eta \cdot \left ( \frac{\Delta \theta _{1} }{\theta _{1} }  \right ) \cdot \mathrm {F}   _{0} \cdot \frac{\sin \pi \mu \Delta \theta _{1} }{\pi \mu \Delta \theta _{1}} exp\left ( -i 2\pi \mu \mathrm {P}   _{\mu }   \right )   ,
\end{equation}

where $\Delta \mathrm {F}\left ( \mu  \right )$ is the variation in the output visibility function, $\mathrm {F}  _{0}  $ is the flux density of the quiet solar profile, $\mu $ is the spatial frequency, $\theta _{1} $ is the angular width of the solar profile, $\Delta \theta _{1} $ is the angular width of the obscured cloud, $\Delta \eta $ is the variation in the cloud transparency, and $\mathrm {P}   _{\mu } $ is the phase difference of the two received signals.

The study is conducted at a frequency of 40 GHz, where the quiet solar radiation flux  $\mathrm {F}   _{0}  $ is approximately 3000 sfu. The results are derived from Formula 4, which indicates that the fluctuations caused by cloud cover with two-element interference are $ \Delta F \sim $ 60 sfu, while without the use of a two-element interference system, the fluctuations are  $\Delta \mathrm {F}=\Delta \eta \cdot \left ( \frac{\Delta \theta _{1} }{\theta _{1} }  \right ) \cdot \mathrm {F}   _{0} \sim$ 600 sfu.The difference between the two is approximately one order of magnitude, and which can be attributed to the interferometer method's ability to suppress the increase in cloud absorption as $\Delta \theta _{1} $ increases.

\section{The design of the system} \label{sec:design}

The composition of the developed two-element interferometric system for solar radio bursts is shown in Fig. 3. The system consists of an antenna, an analog receiver, a digital correlator, and an industrial control computer. Two antennas receive solar radiation signals and output 39.5 GHz$\sim$40 GHz RF signals. The RF signal outputs a 930 MHz$\pm$250 MHz IF signal through two analog receivers. The IF signal is processed by a digital correlator for correlation and outputs the visibility function amplitude and phase, which is transmitted to the IPC through PCIe.

As shown in Figure 4 for the 39.5 GHz$\sim$40 GHz two-element interferometric system, two 50 cm aperture, left circularly polarized Cassegrain-type antennas mounted on a coplanar sliding table module mount to receive solar radiation signals, each with a beam width of ~70 arcmin, which is much larger than the angular size of the solar disk. The distance between the two antennas can be adjusted smoothly between 100 wavelengths and 300 wavelengths, and the distance between the antennas is fixed at 230 wavelengths during observation to reduce the quiet solar radiation. The system performance parameters are shown in Table 2.

\begin{figure}[htbp]
	\centering
	\begin{minipage}[t]{0.48\textwidth}
		\centering
		\includegraphics[width=7cm]{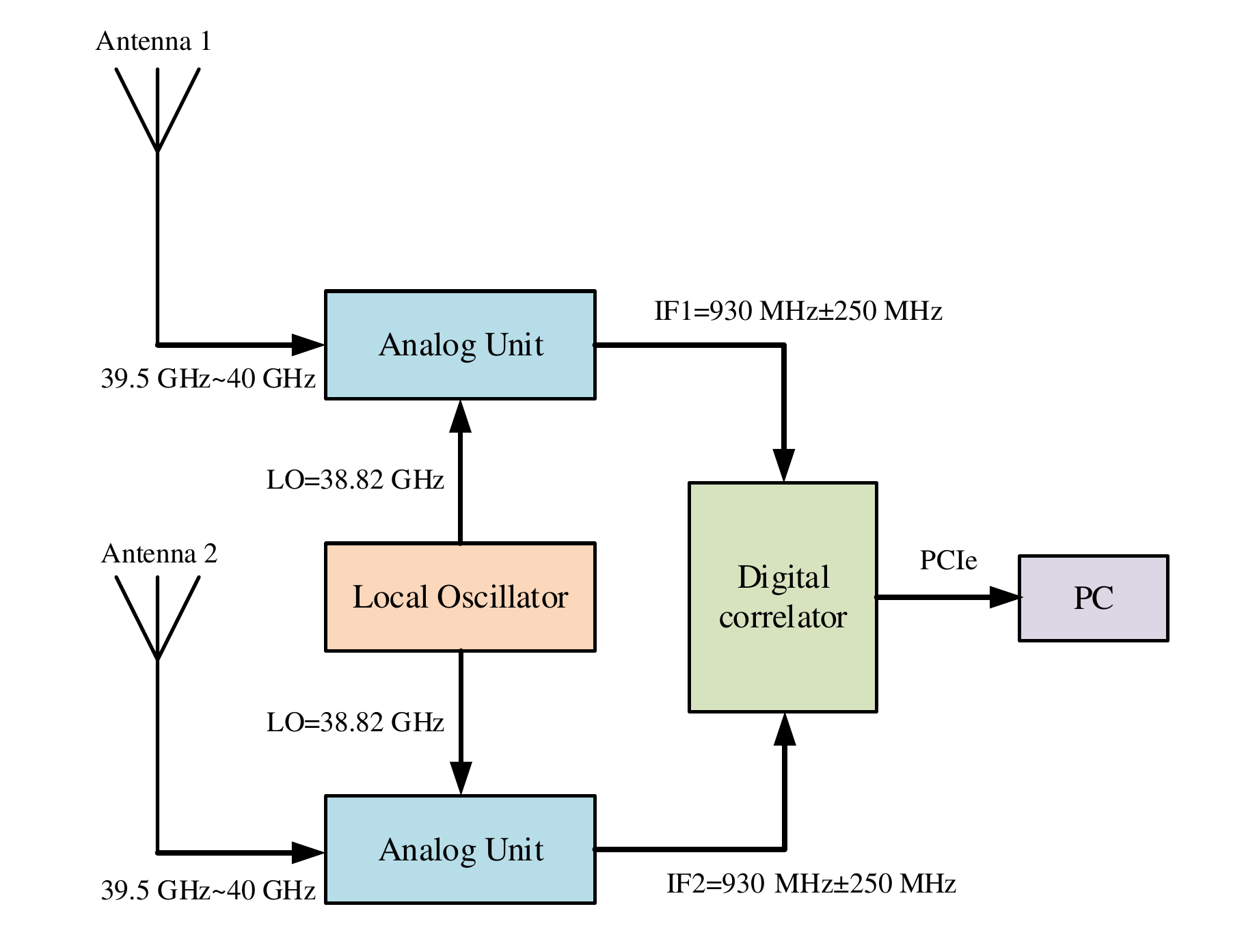}
		\\ \caption{The architecture of the 39.5 GHz$\sim$40 GHz two-element interference system. The system consists of antennas, an analog receiver, a digital correlator, and an IPC \label{fig:general}}
	\end{minipage}
	\begin{minipage}[t]{0.48\textwidth}
		\centering
		\includegraphics[width=5.5cm]{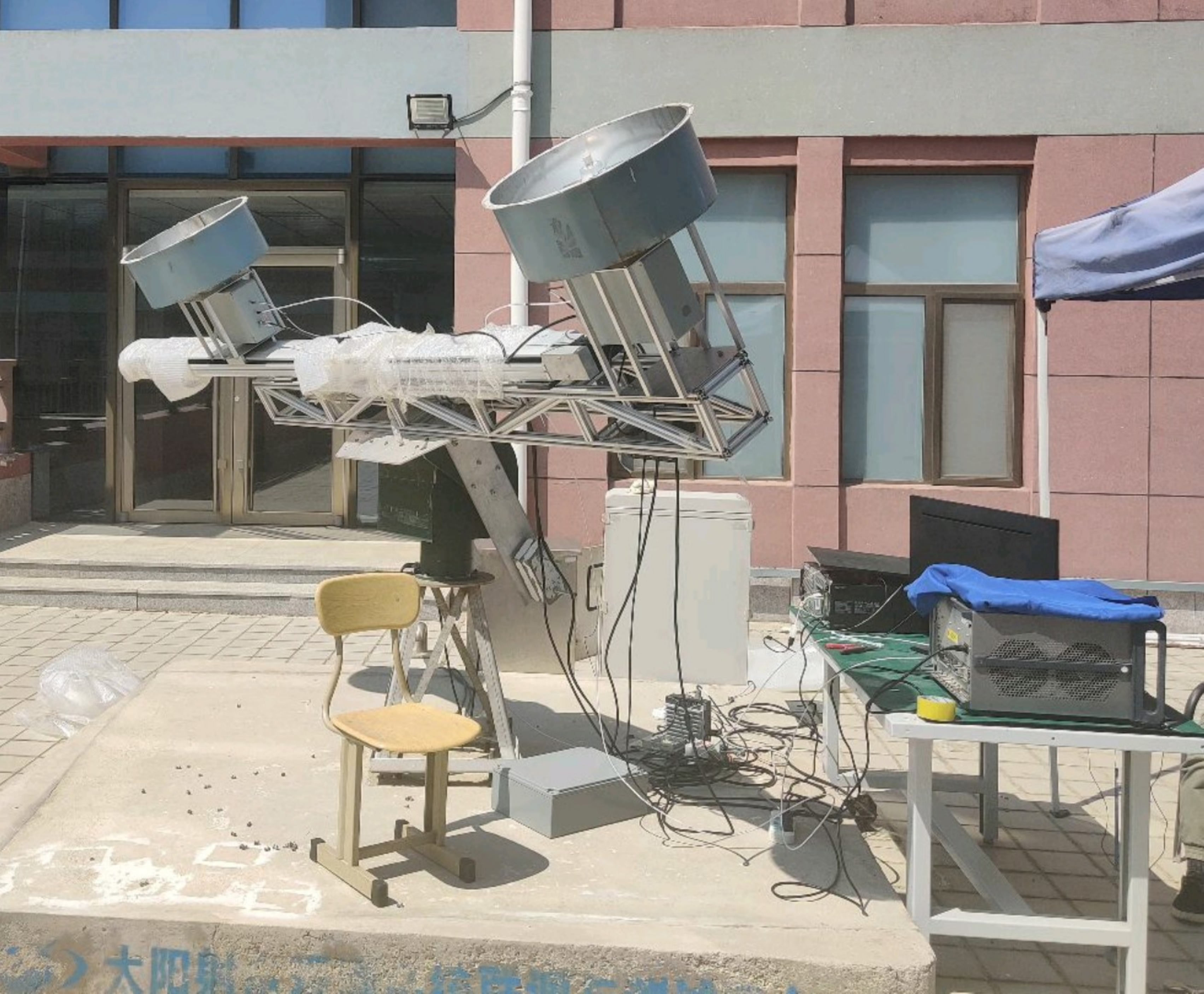}
		\\ \caption{The 39.5 GHz$\sim$40 GHz two-element interferometric system. Two small antennas with a 50 cm diameter (both sides) are mounted on a coplanar slide-out module mount. The distance between the two antennas can be slide-adjusted between 100 and 300 wavelengths  \label{fig:general}}
	\end{minipage}
\end{figure}

\begin{deluxetable*}{cchlDlc}
	\tablenum{2}
	\tablecaption{the main technical specifications of the two-element interferometric system for solar radio observation\label{tab:messier}}
	\tablewidth{0pt}
	\tablehead{ \colhead{Parameter} & \colhead{Value} }

	\startdata
	Frequency & $39.5\thinspace GHz\sim 40\thinspace GHz$ \\
	Frequency Resolution & $153\thinspace kHz$ \\
	Time Resolution & $0.1\thinspace ms$ \\
	Antenna Diameter & $0.5\thinspace m$\\
	Antenna Gain & $41\thinspace dB$  \\
	Analog Front-end Gain & $61\thinspace dB$  \\
	Noise Figure & $\le2.1\thinspace dB(\sim 180\thinspace K)$\\
	Dynamic Range & 3$0\thinspace dB$\\
	Baseline Length & $230 wavelengths \pm  15 wavelengths$ \\
	Polarization Method & $Left\ Circular\ Polarization$  \\
	Spatial Resolution & $14.95 arcmin$\\
	\enddata
\end{deluxetable*}

\subsection{ The Antenna Unit} \label{subsec:tables}

The antenna type adopted two identical Cassegrain antennas, consisting of a primary reflector, a secondary reflector, and a feed source. The primary reflector is a rotating paraboloid, and the secondary reflector is a rotating hyperboloid (\cite{2017IAWPL..16...99Z}). The phase center of the feed source is located at the real focus of the hyperboloid, and the imaginary focus of the hyperboloid coincides with the focus of the paraboloid. Two identical left circularly polarized Cassegrain-type antennas are selected (as shown in Figure 5), operating at 35 GHz$\sim$40 GHz, with maximum radiation gain in the frequency band above 41 dB, beam width of $1.2^{\circ}$, side flap less than -25 dB (as shown in Figure 5), and beam width completely \textbf{covers} the sun ($0.5^{\circ}$). The antenna is mounted on a slide module with a length of 3 m for baseline adjustment (as shown in Figure 4). The minimum variable distance of the slide module is 0.1 mm, and the baseline accuracy is $\pm$0.05 mm.

The specific test performance is shown in Table 3.

\begin{deluxetable*}{lllrrrrrrll}
	\tablenum{3}
	\tablecaption{The antenna gain and 3dB beam width \label{tab:messier}}
	\tablewidth{0pt}
	\tablehead{
		\colhead{ Frequency} & \colhead{ Gain} & \colhead{ 3\thinspace dB Beam-Width}
	}
	\startdata
	$37.5\thinspace GHz$ & $41.9\thinspace dB$ & \quad \quad\quad$1.20^{\circ}$ \\
	$40.0\thinspace GHz$ & $42.4\thinspace dB$ & \quad \quad\quad$1.12^{\circ}$\\	
\enddata	
\end{deluxetable*}

\begin{figure}[htbp]
	\centering
	\begin{minipage}[t]{0.48\textwidth}
		\centering
		\includegraphics[width=5cm]{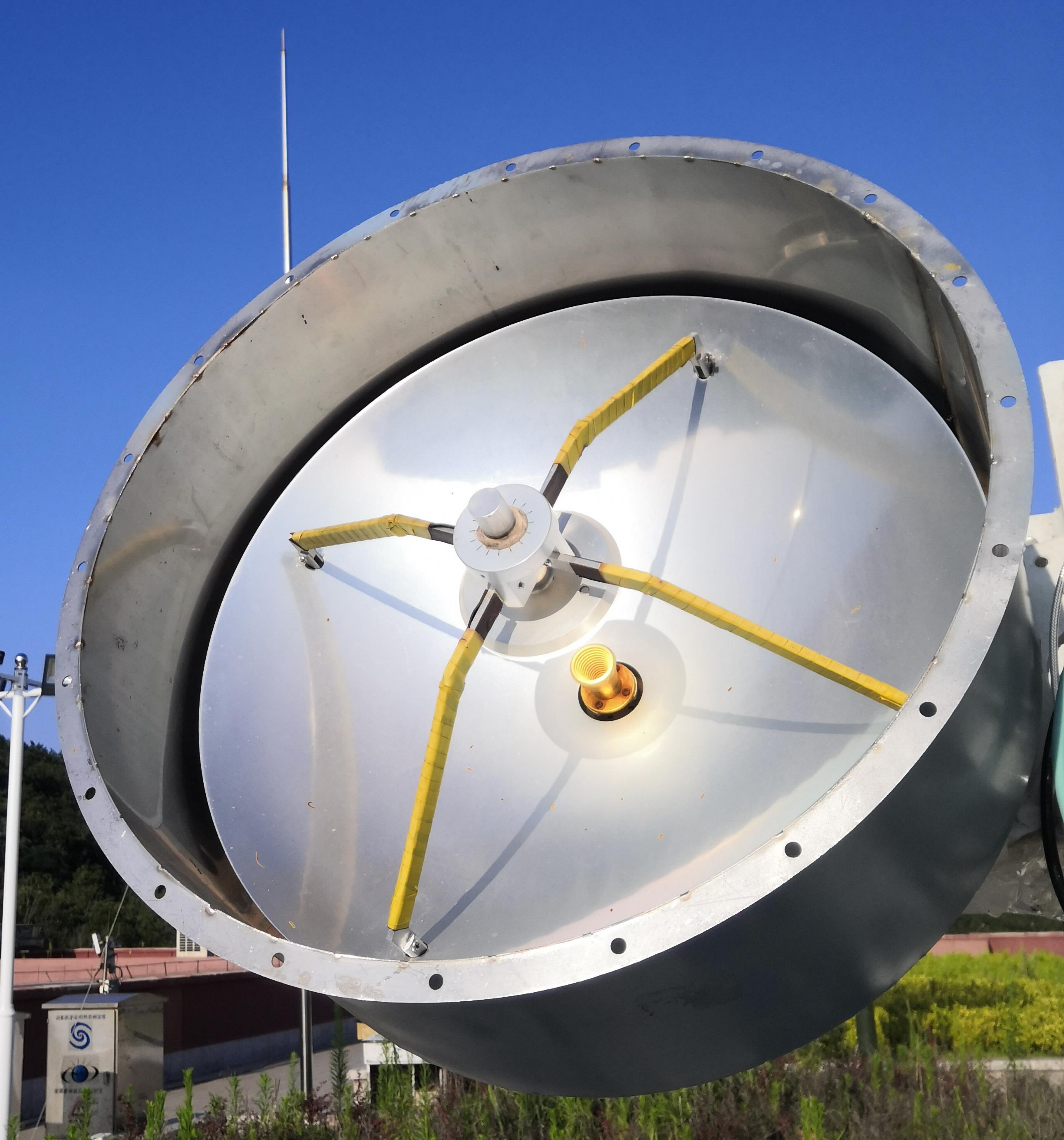}
		\\($a$)
	\end{minipage}
	\begin{minipage}[t]{0.48\textwidth}
		\centering
		\includegraphics[width=7cm]{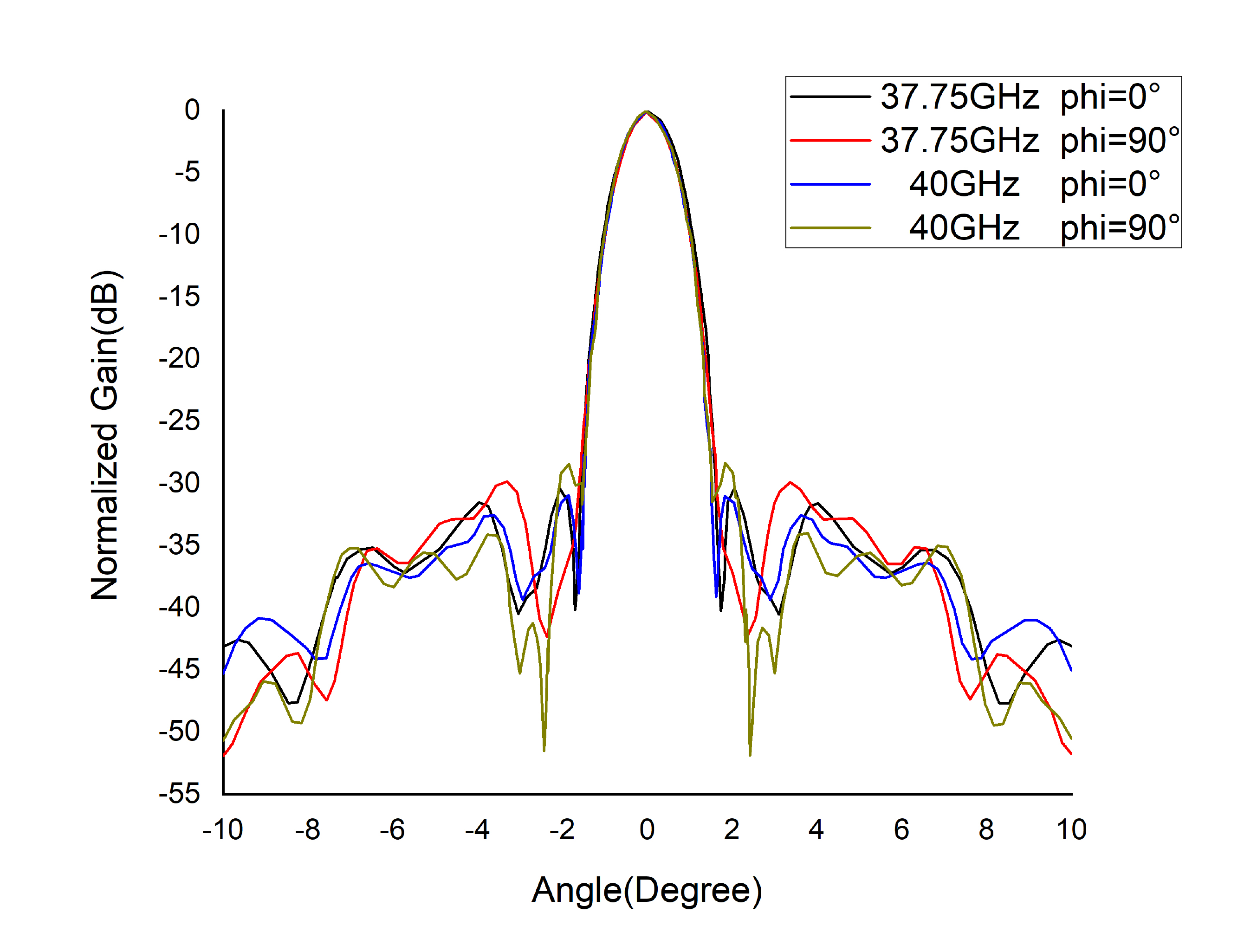}
		\\($b$)
	\end{minipage}
\caption{(a)The physical diagram of the Cassegrain-type antenna;(b)the antenna pattern \label{fig:general}}
\end{figure}

\subsection{ The analog front end} \label{subsec:tables}

The analog front end is responsible for processing the RF signals received from the antenna to generate IF signals, which are tailored to a specific range of frequencies and power levels and subsequently adjusted to suit the requirements of the digital correlator. The analog front-end is shown in Figure 6 with a single-stage mixed super external differential analog receiver. The RF signal output from the antenna is amplified by the low-noise amplifier, and due to the wide amplification band of the low-noise amplifier, filtering is required to filter out the signal outside 39.5 GHz$\sim$40 GHz, and the signal after filtering is input to the mixer for frequency reduction processing. The two analog front ends have a common local oscillator module, which takes into account the sampling frequency of the subsequent ADC module and its power threshold for signal processing, as well as whether crosstalk exists between the local oscillator signal and the RF signal and whether the cross-tuned signal is generated in the IF bandwidth. The input frequency of the local oscillator is 38.82 GHz, and to ensure the phase stability of the two analog front ends, the local oscillator input to the two analog front ends. The length of the phase-stabilized RF cable of the mixer is the same. After mixing, the output IF signal is 680 MHz$\sim$1.18 GHz, which is processed by a two-stage low-noise amplifier and a bandpass filter and then output.

\begin{figure}[htbp]
	\centering
	\begin{minipage}[t]{0.48\textwidth}
		\centering
		\includegraphics[width=9cm]{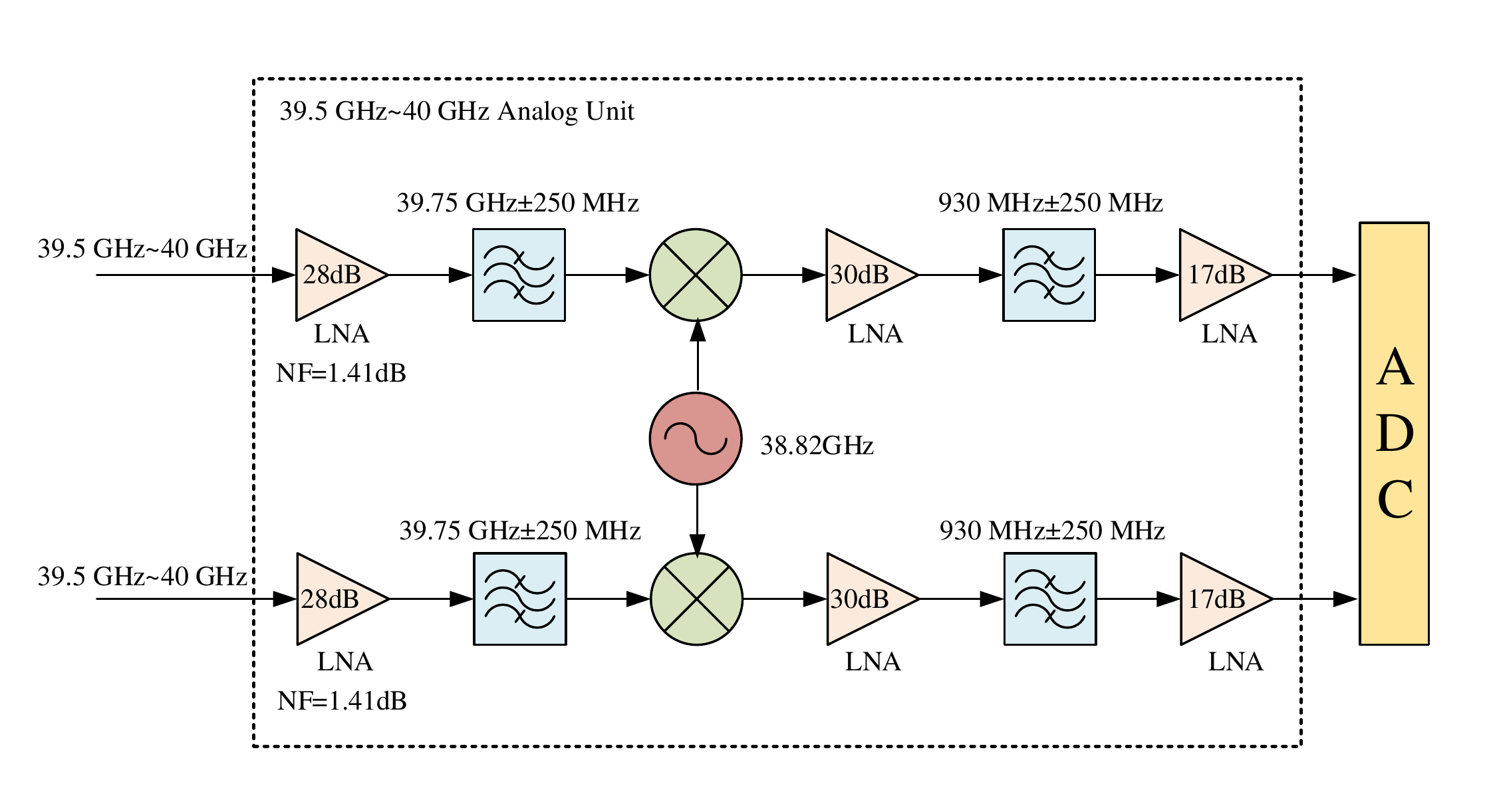}
		\\ \caption{The architecture of the analog front-end block diagram \label{fig:general}}
	\end{minipage}
	\begin{minipage}[t]{0.48\textwidth}
		\centering
		\vspace{-3.7cm}
		\setlength{\abovecaptionskip}{1.0cm}
		\includegraphics[width=10cm]{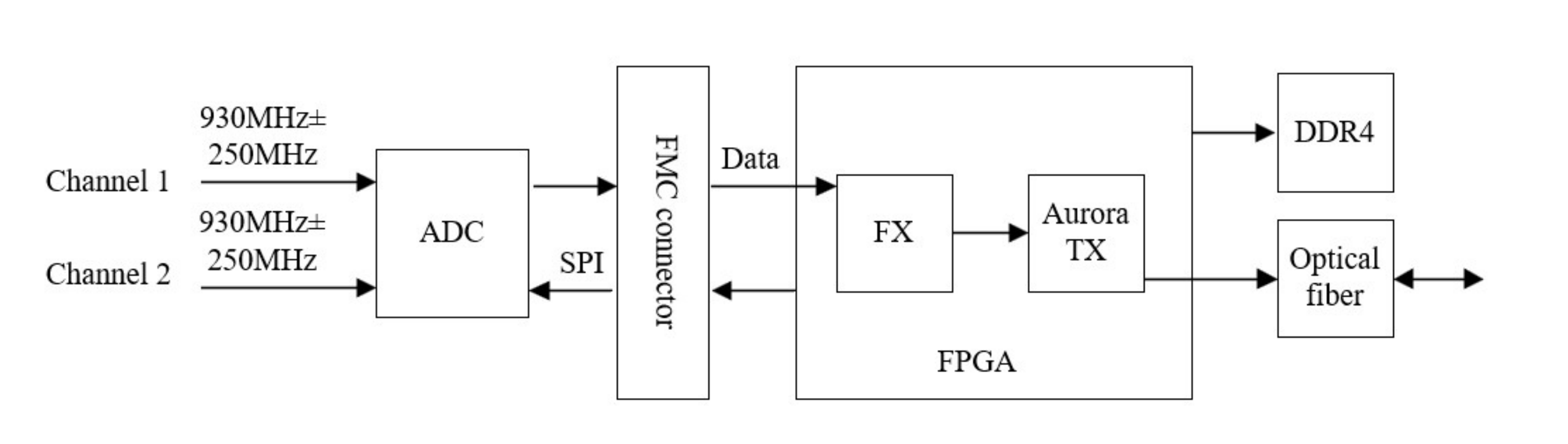}
		\\ \caption{The hardware structure of the digital correlator \label{fig:general}}
	\end{minipage}
\end{figure}

The antenna output interface is a WR$-$28 waveguide port, and the current mainstream practice is to use an adapter to convert WR$-$28 to a coaxial (2.92 mm) interface. Considering that the adapter has a certain insertion loss and the waveguide device has better performance and less noise and insertion loss than the coaxial device, the WR$-$28 waveguide port is used for transmission from the microwave switch to the RF signal input port of the first stage mixer, which will further improve the noise factor of the system.

\subsection{ The Digital Correlator} \label{subsec:tables}

The function of the digital correlator is to sample, digitally quantize and digitally correlate the two IF signals (\cite{2014MNRAS.439.3180F}). The system adopts an FX$-$type digital correlator, as shown in Fig. 7, where the signal data from different antennas are converted to the frequency domain by Fourier transform before the correlation (\cite{1987IEEEP..75.1203C};\cite{2012PASJ...64...29K};\cite{2008PASJ...60..857I}). In these types of correlators, the input bit stream from each antenna is converted into a spectrum by a real-time FFT. Then, for each antenna pair, the spectrum at each frequency is conjugately multiplied to obtain the visibility functions at different frequencies, and the solved visibility functions are summed for a certain time (the summing time is the time resolution).

The digital correlator consists of an ADC acquisition card and an FPGA development board. The ADC acquisition card is AD9691 with a sampling rate of 1.25 Gsps and 14$-$bit conversion accuracy (ADC dynamic range $\ge$ 60 dB, effective bit count 9 bit), which can realize the simultaneous acquisition of two channels (Figure 8). The FPGA development board is equipped with xcku115 series chips, which have rich resources for data processing, and the ADC is connected to the FPGA development board through the FMC physical interface using the JESD204B protocol, which can process high-speed digital data from the ADC. It also has a PCIe$\times$8 socket to support bidirectional data transfer between the board and the host computer. During the actual observation, the acquisition card \textbf{is} independently powered. It receives the data transmitted from the acquisition card and transmits the data to the host computer via PCIe.

\begin{figure}[htbp]
	\centering
	\begin{minipage}[t]{0.48\textwidth}
		\centering
		\includegraphics[width=6cm]{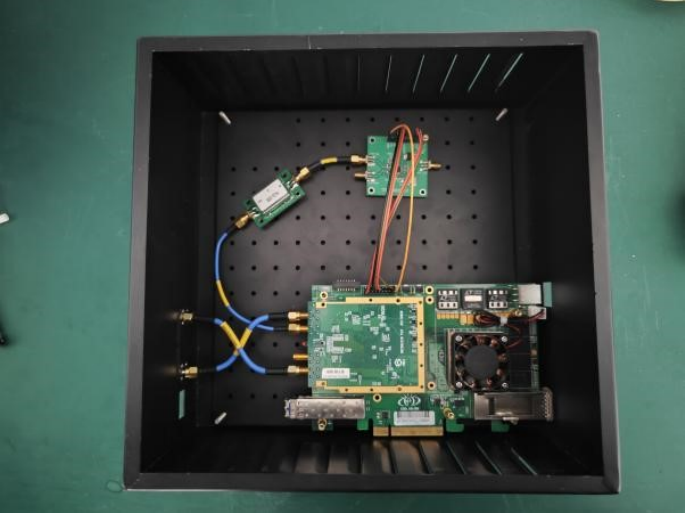}
	\end{minipage}
	\begin{minipage}[t]{0.48\textwidth}
		\centering
		\includegraphics[width=6cm]{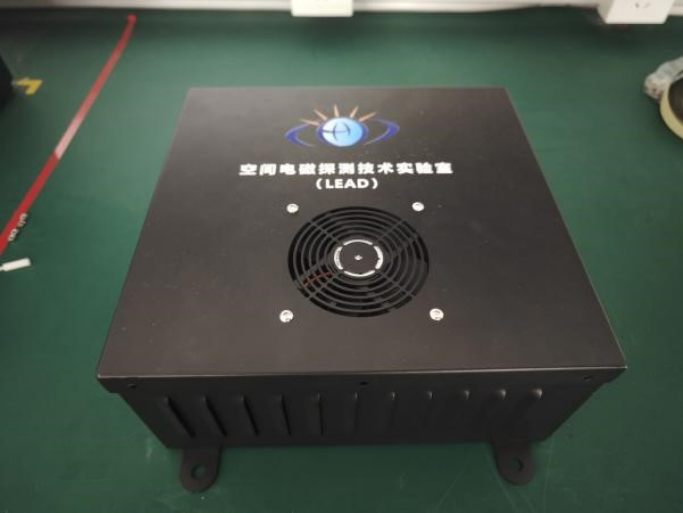}
	\end{minipage}
	\caption{Digital correlator prototype \label{fig:general}}
\end{figure}

\begin{figure}[ht!]	
	\includegraphics[width=100mm]{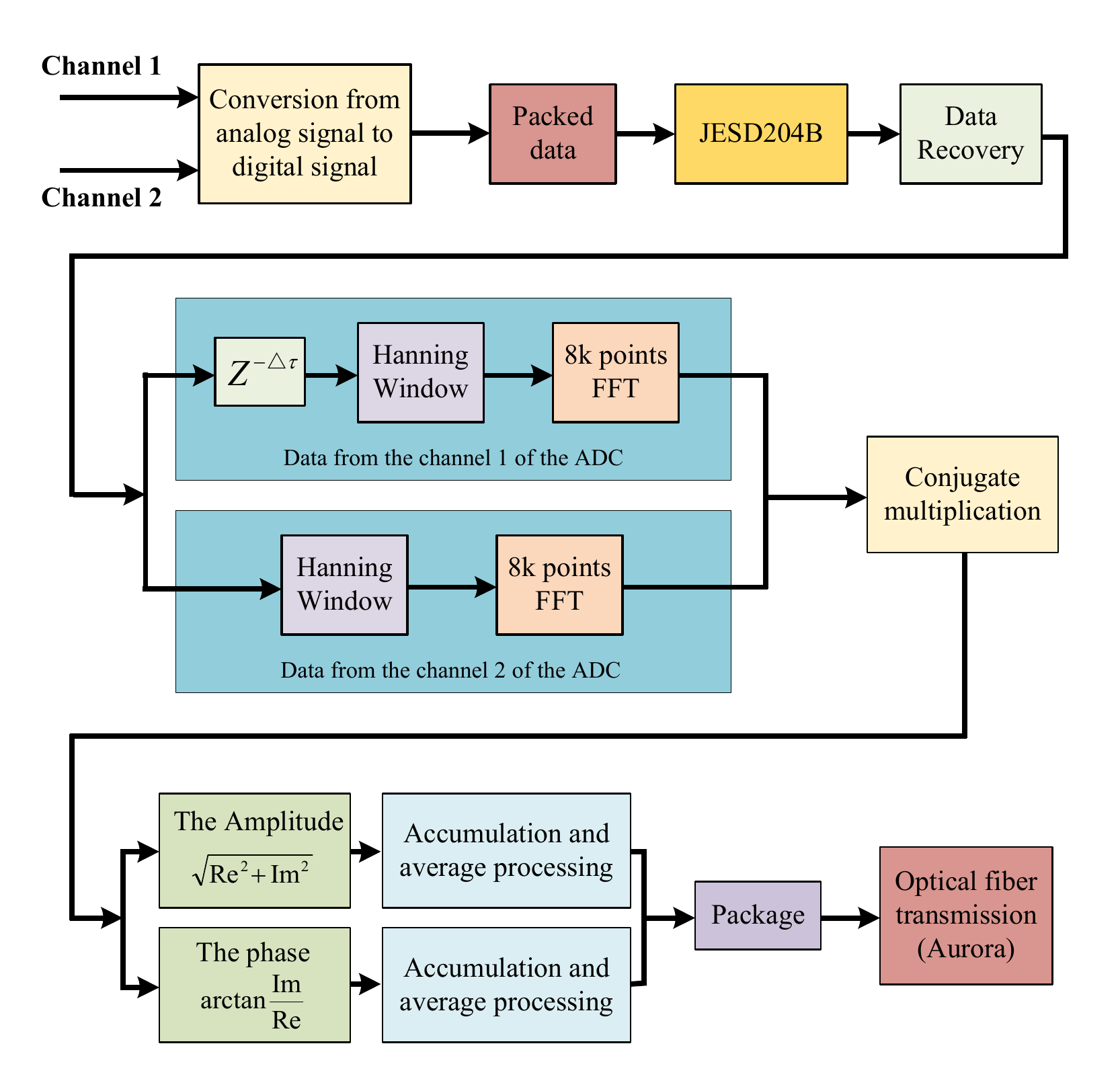}
	\centering
	\caption{The signal processing flow chart of the digital correlator  \label{fig:general}}
\end{figure}

As shown in Fig. 9, the digital correlator transfers the 930 MHz$\pm$250 MHz IF signal from different antennas and analog front ends to the FPGA through the ADC acquisition card with 1.25 Gsps sampling and 14$-$bit quantization and then transfers it to the FPGA through the FMC interface. The ADC data of each channel are divided into 8 channels after entering the FPGA. First, the 8-way data are processed separately by adding windows (hanning windows) to prevent spectrum leakage (\cite{2021PASJ...73..439Y}), and the window function is generated by the MATLAB toolbox, after which 1k$-$point FFT processing is performed simultaneously for each windowed data (the sampling rate is 1.25 Gsps, the FFT points are 8192, and the subchannel bandwidth is 153 kHz; i.e., the frequency resolution is 153 kHz). Then, a three-stage butterfly operation is performed to obtain the frequency domain data of that channel ADC, converting the time domain signal into a frequency domain signal (\cite{2022RAA....22h5012Z}). The phase noise of the analog front-end device and the electrical length of the transmission cable are inconsistent, resulting in a slope line (with slope k) in the phase spectrum. Therefore, after FFT operation in FPGA, one channel is used as the reference, and then each frequency point of the other data is multiplied by the phase shift factor  $e^{-2\pi v_{IF} \Delta \tau } $. The coefficient varies for different frequency channels. The visibility function is obtained by conjugate multiplication of the two frequency domain signals, the amplitude and phase of the visibility function are obtained, the power spectrum and phase spectrum are accumulated afterward, and the number of accumulations is adjustable.

The data processed by the digital correlation receiver are the power and phase of the visibility function, and the data are grouped into frames. Each frame is 128 KB in size, and each frame is 1025$\times$4 points, where the first point of each group of 1025 points is the frame header data. The upper computer reads these data for graphing.

The interaction between the receiver and the host computer is realized through PCIe. The receiver board comes with a PCIe $\times$ 8 interface, and the development board is inserted into the PCIe slot of the host computer to realize bidirectional data transfer. The receiver board transmits the raw data to the host computer, which saves the raw data to the disk array. The digital correlator performance parameter indicators are shown in Table 4.

\begin{deluxetable*}{cchlDlc}
	\tablenum{4}
	\tablecaption{The performance parameters of the digital correlator specification\label{tab:messier}}
	\tablewidth{0pt}
	\tablehead{
		\colhead{Parameter} & \colhead{Value}
	}
	\startdata
	Correlation Type & $FX$ \\
	ADC Sampling Rate & $1.25\thinspace Gsps,14\thinspace bit$ \\
	ADC Effective Bits & $9\thinspace bit$ \\
	ADC Dynamic Range & $\ge 60\thinspace dB$\\
	IF Input Frequency & $680\thinspace MHz \sim 1180\thinspace MHz$   \\
	Frequency Resolution & $153\thinspace kHz$ \\	
	Time Resolution & $0.1\thinspace ms\left ( Default \right )$ \\
	Digital Signal Transmission Interface & $SFP+  \left ( Board \ to\ board\ communication \right)   /PCIe  \left (Board\ to\ host\ computer \right)$  \\
	Delay Range & $0 \sim 8\thinspace ns$ \\	
	Delay Step & $0.8\thinspace ns \quad \left ( 1/1.25\thinspace GHz \right)$  \\		
	\enddata
\end{deluxetable*}

\section{System performance} \label{sec:performance}

The development of the prototype of the two-element interferometer has been completed, In the following, the performance of the system is presented in terms of the gain, noise figure, input-output characteristics, and observations obtained through system testing and data analysis.

\subsection{ The analog front end} \label{subsec:tables}

The lower the noise factor of the system is, the higher the sensitivity of its system, and to observe small flare bursts as much as possible, the noise factor of the system should be as low as possible. The noise factor of the system is given by the following equation (\cite{2004Noise...51..1330N}).

\begin{equation}
N  _{rec} =N_{1}+  \left [ \frac{N_{2}-1 }{G_{1} }  \right ]  +  \left [ \frac{N_{3}-1 }{G_{1}G_{2}  }  \right ] + \cdots   ,
\end{equation}

where $N  _{rec}$ is the system noise factor; $N_{1}$, $N_{2}$, and $N_{3}$ are the noise factors of the first-stage amplifier, second-stage amplifier, and third-stage amplifier, respectively; and $G_{1}$ and $G_{2}$ are the corresponding gains of the first-stage amplifier and second-stage amplifier, respectively. As seen from Equation (5), the value of the system noise factor is mainly determined by the noise factor and gain of the first-stage amplifier, so the selection of the first-stage device with a low noise factor and large gain can effectively reduce the noise factor. The first-stage low-noise amplifier selected for use has a WR$-$28 waveguide port at both ends and a noise figure of 1.41 dB (111.2 K) in the 39.5GHz$\sim$40 GHz band at room temperature, with a gain of 27 dB. The noise figure of the two analog front ends is below 2.1 dB (180.3 K), as shown in Figure 10.

The two analog front-ends need to have a certain gain to ensure that the ADC can detect the radiation power of the quiet sun and that the dynamic range of the system is as large as possible. The radiated power received by a single link can be estimated from Equation (6).

\begin{equation}
P_{sys} =k\left ( T_{a}+  T_{rec}   \right ) B G_{sys}   ,
\end{equation}

where $T_{a}$ is the antenna temperature, $T_{rec}$ is the receiver noise temperature, $B$ is the observation bandwidth, and $G_{rec}$ is the front-end gain. $T_{a}$ and $T_{rec}$ can be obtained from the following equation (\cite{2017isra.book.....T}).

\begin{equation}
	T_{a} =\frac{A_{e}S_{sun}  }{2k}    ,
\end{equation}

\begin{equation}
	T_{rec} =\left (10 ^{N_{rec}/10 } -1 \right ) T   ,
\end{equation}

where $A_{e}$ is the effective area of the antenna, $S_{sun}$ is the quiet solar radiation flux density corresponding to the observed frequency, $k$ is the Boltzmann constant, $N_{rec}$ is the system noise factor, and T is 290 K.

\begin{equation}
	A_{e} =\frac{G_{a}\lambda ^{2}  }{4\pi }  ,
\end{equation}

where $G_{a}$ is the antenna gain and $\lambda$ is the wavelength corresponding to the observed frequency.

\begin{figure}[htbp]
	\centering
	\begin{minipage}[t]{0.48\textwidth}
		\centering
		\includegraphics[width=6.5cm]{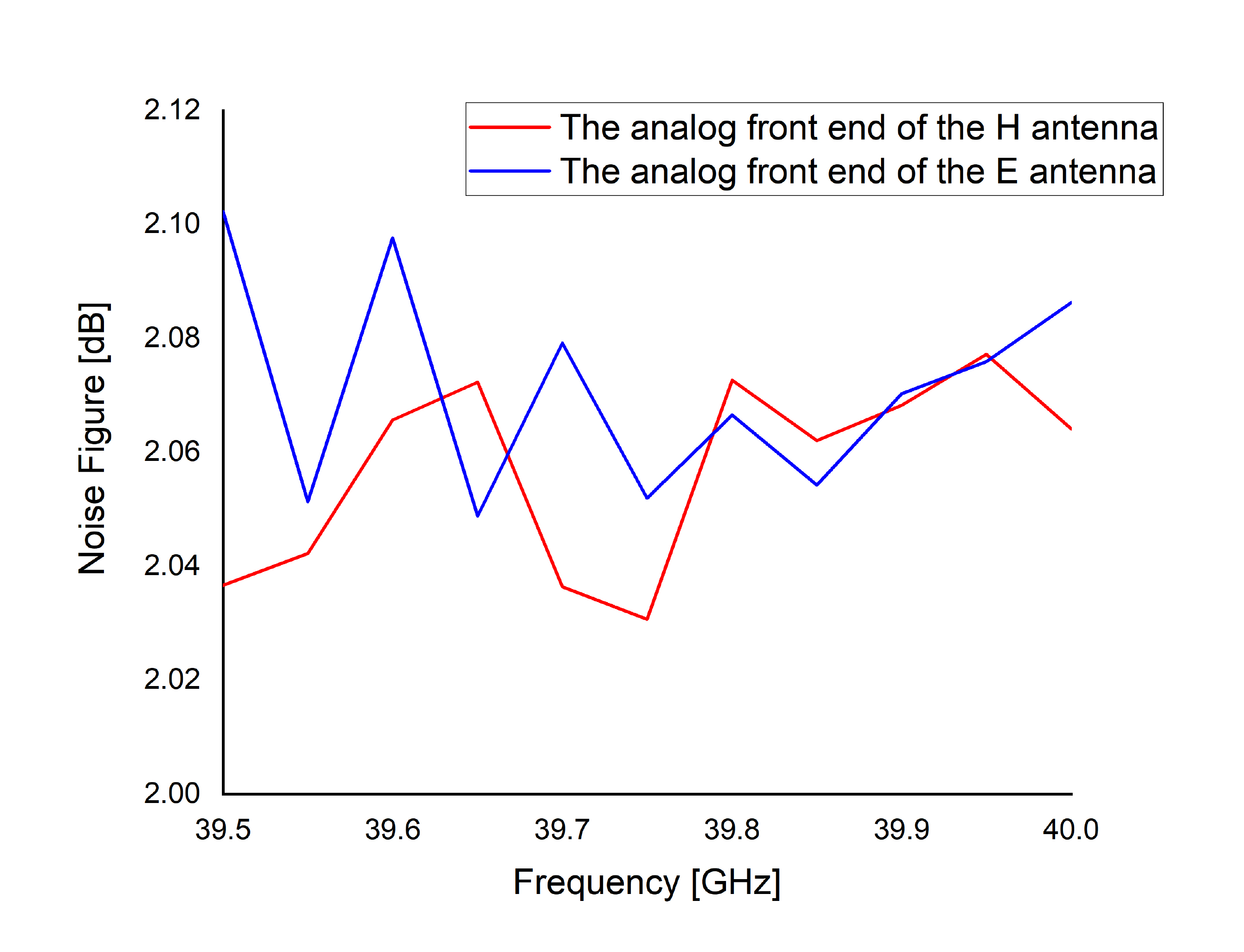}
		\\ \caption{H and E analog front-end noise measurements \label{fig:general}}
	\end{minipage}
	\begin{minipage}[t]{0.48\textwidth}
		\centering
		\includegraphics[width=6.5cm]{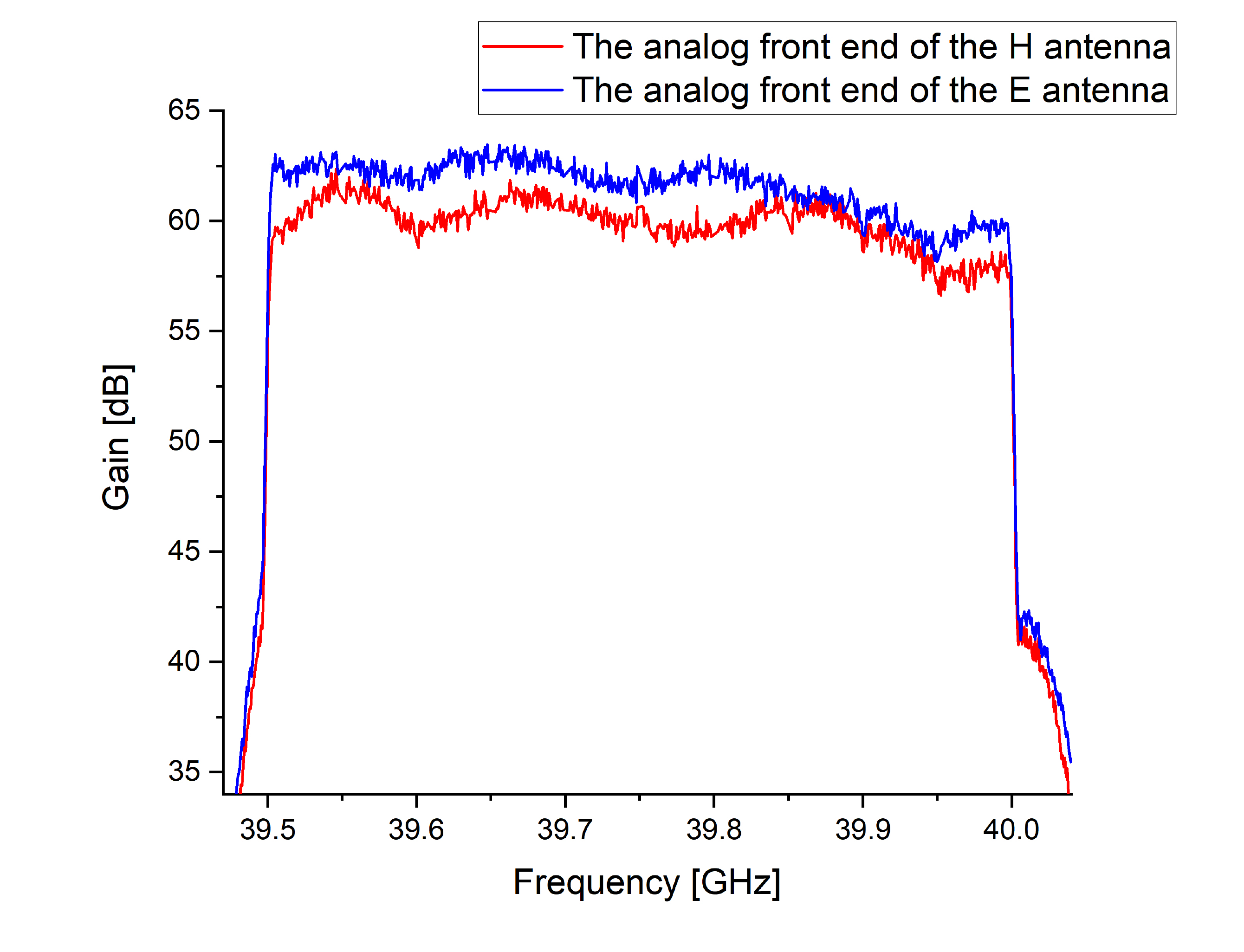}
		\\ \caption{ H and E analog front-end gain comparison  \label{fig:general}}
	\end{minipage}
\end{figure}

The noise floor of the quiet sun in the 39.5 GHz$\sim$40 GHz band is $-$164.4 dBm/Hz, and the ADC minimally detects the signal of $-$107 dBm/Hz. Therefore, to allow the system to detect the quiet sun signal, the effective gain of the two analog front ends must be at least 57.4 dB, and a suitable system gain will provides a suitable dynamic range for solar activity. The effective gain of this system is approximately 61 dB, as shown in Figure 11.

\subsection{ Interferometer output compared with single output} \label{subsec:tables}

From Fig. 1, it can be seen that the amplitude and phase are obtained when the two-element interferometric system observes solar radiation. The baseline of the system is fixed at $230\lambda$ for the observation of the Sun, as shown in Figure 12. The output amplitude between 39.5 GHz and 40 GHz in the observation band is significantly elevated, indicating that the system has received the solar radiation signal, whose amplitude is flatter in the band. The phase in the corresponding frequency band fluctuates near 0 degrees (-5 degrees to +5 degrees), indicating that the solar radiation signal basically arrives at the digital correlator input at the same time. Therefore, from the data results in Figure 13, it is clear that the two-element interferometric system can simultaneously receive the radiation signal from the quiet solar profile.

\begin{figure}[htbp]
	\centering
	\begin{minipage}[t]{0.48\textwidth}
		\centering
		\includegraphics[width=6.5cm]{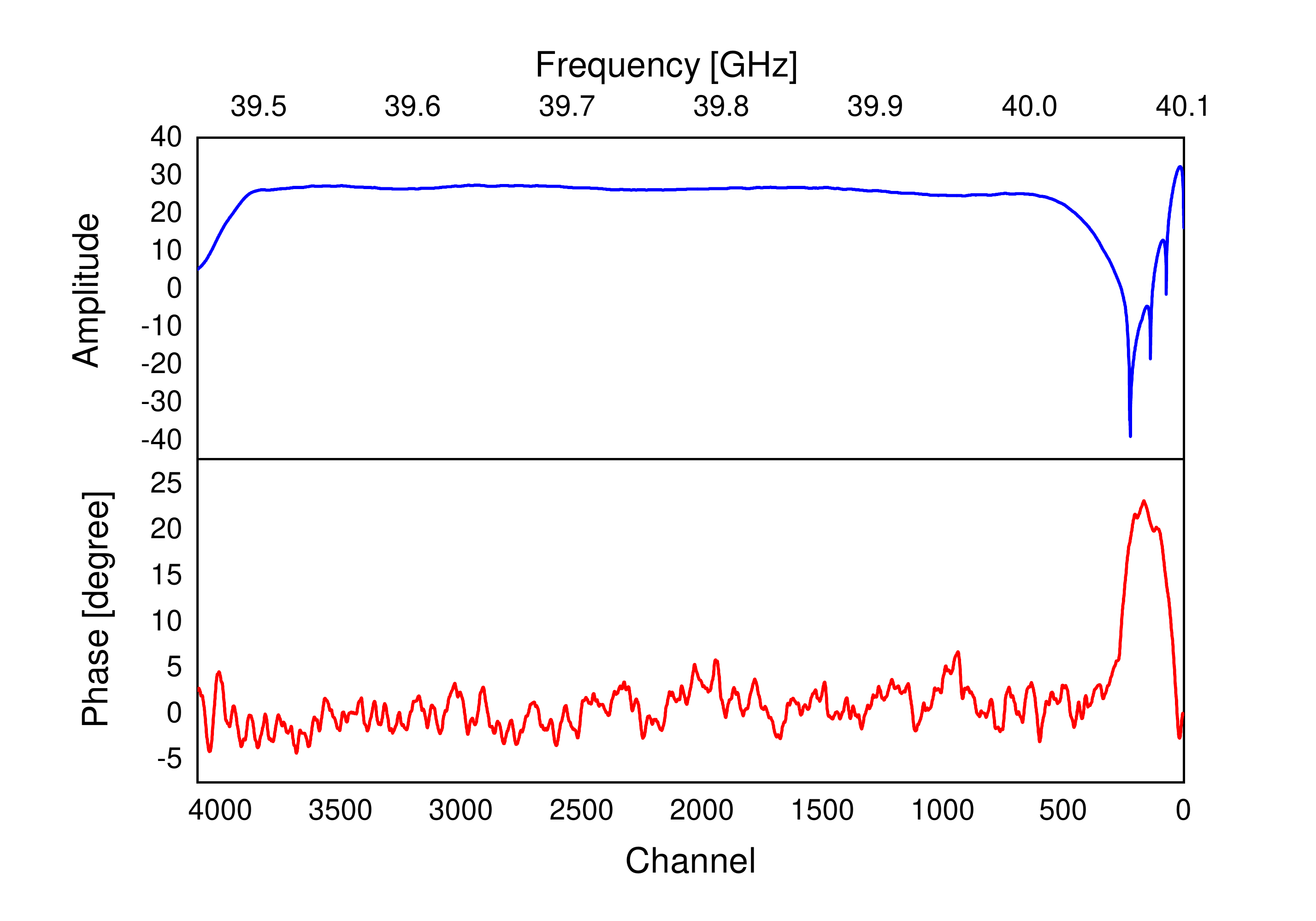}
		\\ \caption{The amplitude and phase of the interferometer output in the observation band when observing the solar \label{fig:general}}
	\end{minipage}
	\begin{minipage}[t]{0.48\textwidth}
		\centering
		\includegraphics[width=6.5cm]{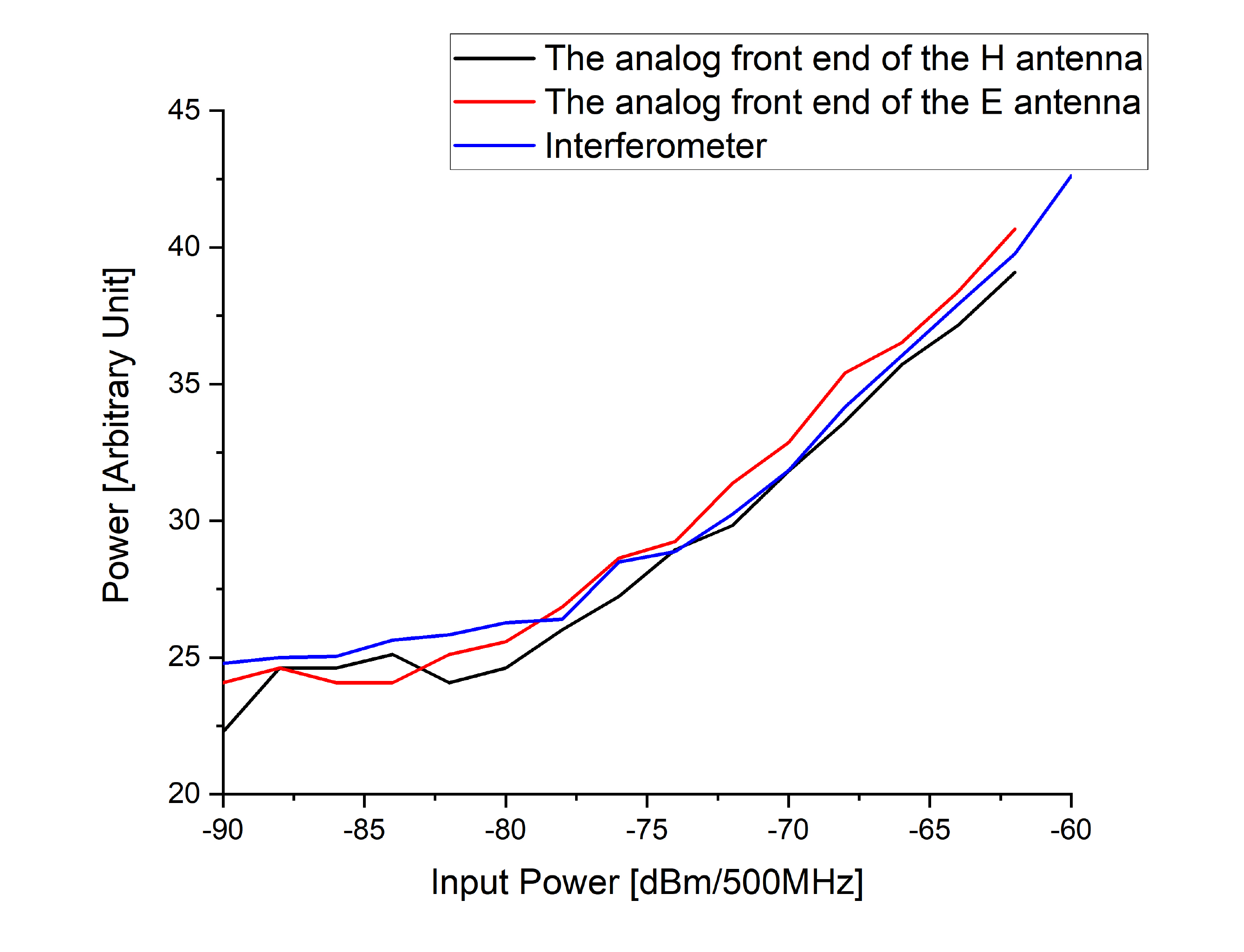}
		\\ \caption{ The interferometer output compared with the H and E single antenna output  \label{fig:general}}
	\end{minipage}
\end{figure}

The amplitude of the digital correlator output to observe the sun (\cite{1998ITGRS..36..822L}) is $F=\sqrt{R_{C}^{2}+ R_{S}^{2} } $, where $R_{C}\sim A\left ( \theta  \right ) \cos \left ( \varphi  \right ) $ and $R_{S}\sim A\left ( \theta  \right ) \sin \left ( \varphi  \right ) $. Therefore, the amplitude of the interferometer output is proportional to the antenna pattern $A\left ( \theta  \right )$ and corresponds to the mutual power spectrum of the signals received by the two antennas. Since the antenna receives and amplifies the solar radiation signal, and then the analog front-end amplifies it to the IF signal captured by the adaptive ADC, the amplitude of the correlator output should be

\begin{equation}
	F=\sqrt{W_{H} G_{H}W_{E}G_{E}} \cos \left ( \varphi  \right )   ,
\end{equation}

where $W$ is the radiated power from the sun collected by the antenna, $G$ is the effective gain of the analog front end, and the subscripts H and E denote the two antennas. As shown in Figure 13,  $\varphi \in \left [ - 5^{\circ},+5^{\circ} \right ] $, so $\cos \left ( \varphi  \right ) \approx 1$. The amplitude of the correlator output can be approximated as $	F=\sqrt{W_{H} G_{H}W_{E}G_{E}}$, and the amplitude of the two-element interferometer output should be approximated as the geometric mean of the output amplitude of the two independent antennas. When the directional maps of the two antennas and the parameters of the simulated front end are the same, the amplitude of the two-element interferometer output is the same as the amplitude of the two independent antenna outputs (\cite{1997MNRAS.286...48M}).

Using the vector signal source output 39.75 GHz$\pm$250 MHz bandwidth signal, through the analog front-end input to the two-element interferometer, there is a signal input power range of $-$90 dBm to $-$60 dBm, while analyzing and comparing the two-element interferometer output power and H, E single output power. As shown in Figure 13, the two-element interferometer output power is basically between the H and E single output power. At low power input, the two-element interferometer and the H and E single antennas enter the nonlinear region, and the single antenna output power fluctuation becomes larger. Moreover, the interferometer output is slightly flatter than that of the single, so the interferometer is more able to identify small signals such as small solar flare outburst signals.

\subsection{ Preliminary observation results} \label{subsec:tables}

When the system is actually observed, in addition to fluctuations in the observed data due to factors such as system gain fluctuations and noise figures, environmental factors can also cause fluctuations in the observed data to varying levels, such as weather conditions (rain, fog, etc. ), cloud thickness, etc. (\cite{1994MNRAS.268..299W}). Therefore, the response of the test system in clear weather and the comparison of the fluctuation of the single antenna and the interferometric output in the case of thick cloud occlusion are tested.

In clear and cloudless weather, the output of the interferometric system is compared by pointing at the cold sky and the sun, respectively,and the spectrogram when the system is pointing at the sun,  as shown in Figure 14. The radiation flux pointing to the sun at each frequency point in the observation band is approximately 3000 SFU more than that pointing to the cold sky, and the fluctuation range decreases, which also corresponds to the theoretical value of the solar radiation flux at 40 GHz. We can also see that when the interferometer is aimed at the sun, the fluctuation of the radiation flux at each frequency point is smaller and basically controlled within 50 SFU.

\begin{figure}[htbp]
	\centering
	\begin{minipage}[t]{0.48\textwidth}
		\centering
		\includegraphics[width=8.5cm]{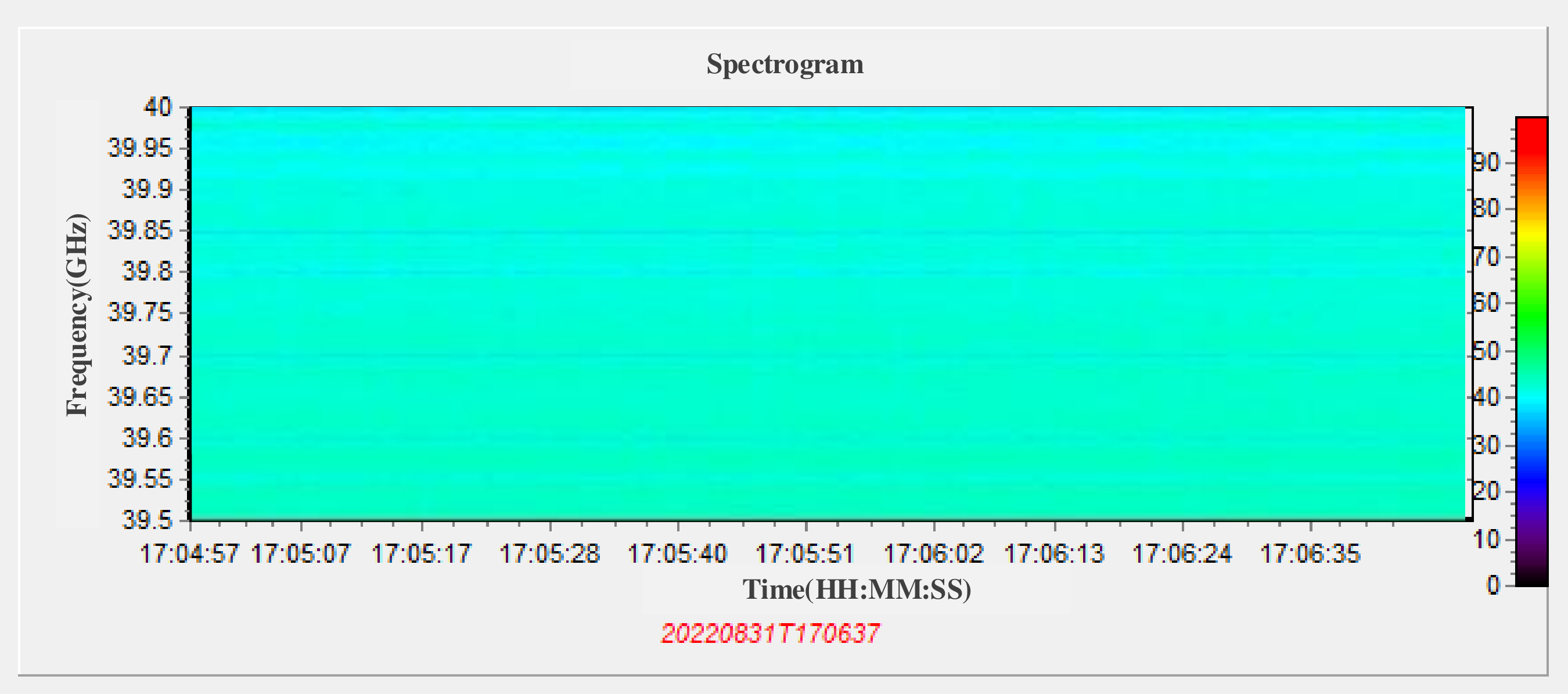}
		\\($a$)
	\end{minipage}
	\begin{minipage}[t]{0.48\textwidth}
		\centering
		\vspace{-10.7em}
		\setlength{\abovecaptionskip}{0.0cm}
		\includegraphics[width=9.1cm]{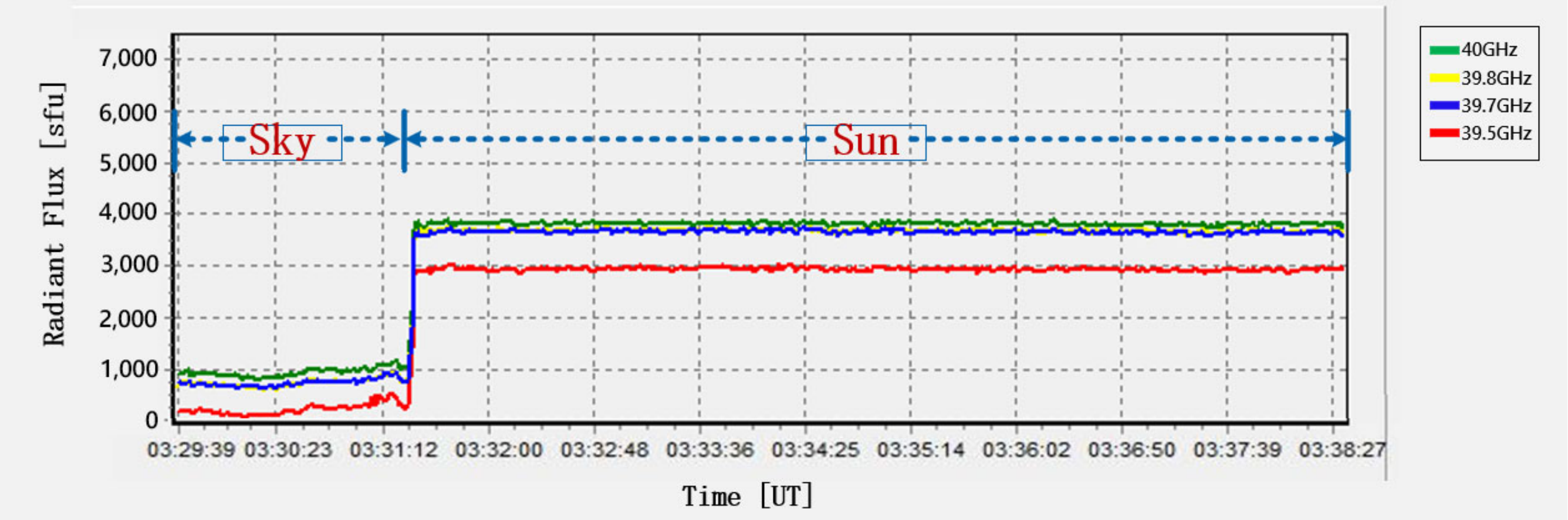}
		\\$ $
		\\($b$)
	\end{minipage}
	\caption{(a) Dynamic spectrum with 500 MHz bandwidth when the system is pointing to the sun; (b) Comparison of the output of the interfering system point to the cold air and the solar. The radiation flux of each frequency pointing to the sun is approximately 3000 sfu more than that when pointing to the cold sky, while the fluctuation of the radiation flux of each frequency point is basically controlled within 50 sfu when aimed at the sun \label{fig:general}}
\end{figure}

Weather with many clouds and thick clouds in the sky is selected for observation, and the interferometric system tracks the solar. The interferometric system outputs the interferometric results in the 40 GHz band while outputting the radiation flux received by a single antenna at 40 GHz, as shown in Figure 15. When cloud occlusion occurs, the radiation flux received by a single antenna at 40 GHz fluctuates more, approximately 900 SFU, and the output fluctuation of the interferometric system is approximately 190 SFU.

\begin{figure}[ht]	
	\includegraphics [width=80mm]{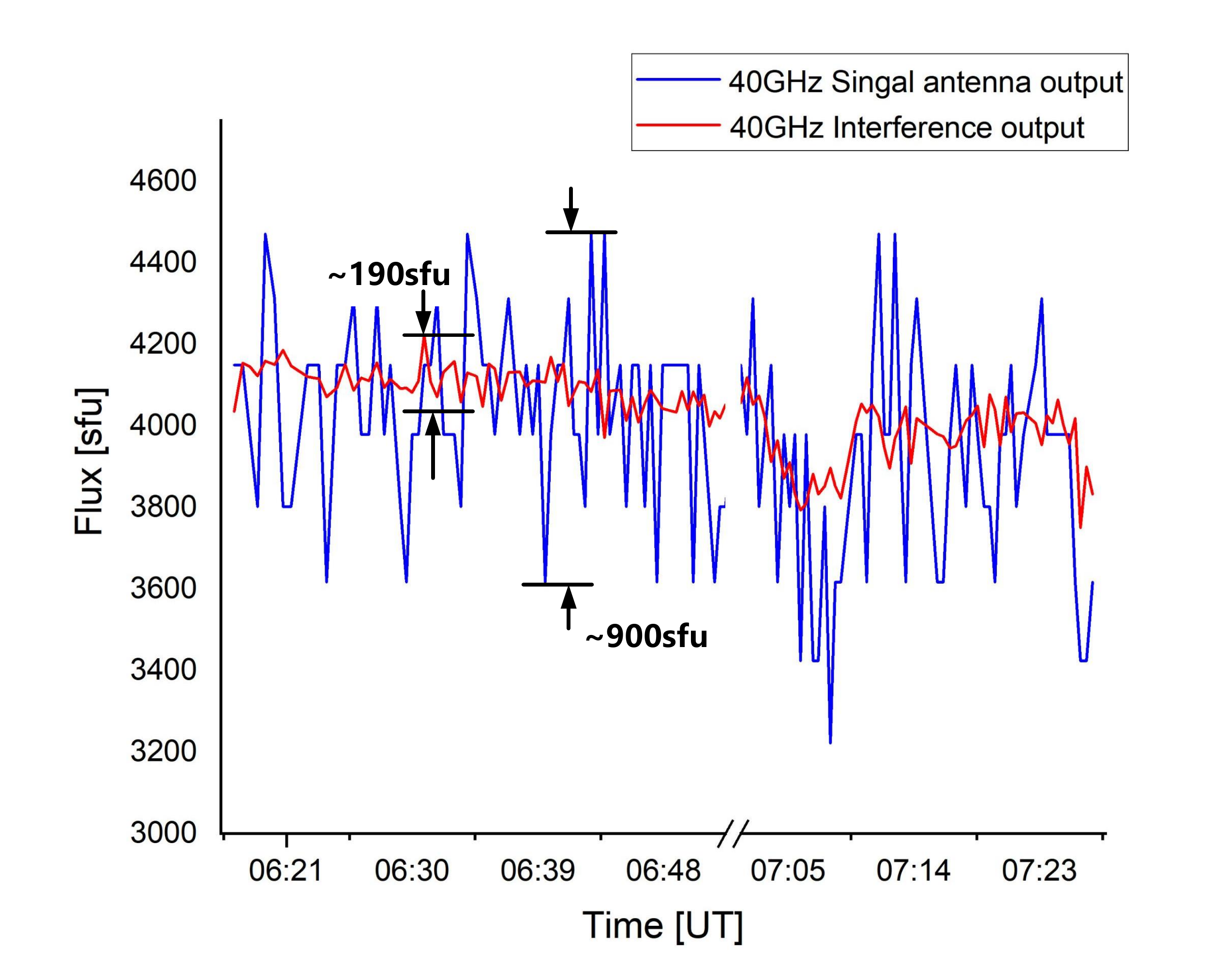}
	\centering
	\caption{Comparison of single antenna and interferometric system output when observing 40 GHz solar signal under a cloudy sky and thick cloud cover. When the cloud is blocked, the radiation flux received by a single antenna at 40 GHz fluctuates by approximately 900 SFU, while the output of the interferometric system fluctuates by approximately 190 SFU. When the thick cloud is blocked, the reduced radiation flux of the single antenna is much larger than that of the interferometric system, and the radiation flux of the interferometric system changes by approximately 300 SFU, while that of the single antenna decreases by approximately 1000 SFU  \label{fig:general}}
\end{figure}

In Fig. 15, we can also see that between $07:06$ and $07:12$,the output of both decreases due to thick cloud occlusion, but the reduced radiation flux of the single antenna output is much larger than that of the interferometric system. The radiation flux of the interferometric system changes by approximately 300 SFU, while that of the single antenna decreases by approximately 1000 SFU. In comparison, the interferometric system greatly reduces the fluctuation range.

\section{Calibration} \label{sec:Calibration}

In this system, the calibration is the calibration of the amplitude and phase of the entire receiver chain (\cite{2016AN....337.1099V}), which is performed by "internal calibration" and "external calibration".

The vector signal source generates 39.5GHz$\sim$40 GHz calibration signals of different powers, which are input from the waveguide coupler to the receiving link. The "visibility function" of the calibration signal can be obtained through the processing of the calibration signal by two analog front ends and digital correlators (the phase of the visibility function is the phase difference between different chains). Based on this phase, calibration compensation is performed, and this method is called "internal calibration". The internal calibration starts from the calibration of the waveguide coupler and does not include the calibration of the antenna and the feed source. The "internal calibration" method compensates the phase in the observed frequency band to essentially 0 degrees, thus ensuring the consistency of the phase response of the analog front-end.

To complete the calibration of the amplitude between the two antennas and the feed source, the cold sky and the quiet sun are used as natural calibration sources to calibrate the amplitude of the whole link and further improve the calibration accuracy (\cite{2013EM&S....49...118B}). After "internal calibration", the cold sky and quiet solar observations are conducted separately at a certain time, the visibility functions of the cold sky and quiet solar output are obtained, and the difference between the visibility functions is compared with the theoretical calculation. This difference between these is considered to be caused by the difference between the antennas, and this difference is solved to further realize link compensation. This method is called "external calibration". According to the 35 GHz$\sim$40 GHz solar radio dynamic spectrum observation system of the Chashan Solar Radio Observatory (CSO) and the theoretical radiation temperature of the Sun in the 40 GHz band, the radiation flux of the Sun in the 40 GHz band is calculated to be approximately 3000 SFU(\cite{2023ApJ...942L..11Y}). As shown in Figure 14, the difference in the amplitude of the visibility function obtained from the two-element interferometer pointing at the cold sky and the quiet sun linearly corresponds to the radiation flux of the sun in the 40 GHz band.

\section{Conclusion and Discussion} \label{sec:Conclusion}

To improve the system sensitivity, reduce the fluctuation of the whole solar radiation, and realize the observation of solar small flare bursts in the 40 GHz band, a two-element interferometer for solar radio bursts operating at millimeter wavelengths has been developed based on the nulling interference method. It has been installed at the Chashan Solar Radio Observatory (CSO), where it was tested for solar radio observation. The solar radiation signal is received by two 50 cm diameter, circularly polarized Cassegrain-type antennas mounted on the same planar mount at a baseline distance of 230 wavelengths. The nulling interferometer cancels out the radiation from the quiet sun, largely reducing fluctuations in the quiet solar flux density When flares smaller than the entire solar disk occur, the system can detect small solar eruptions with high sensitivity. The performance parameters of the two-element interferometer are as follows: the observation band is 39.5 GHz$\sim$40 GHz, the frequency resolution is 153 kHz, the time resolution is 0.1 ms, the analog front-end noise figures are $\le$2.1 dB, and the system noise temperature is 2400 K.

The system was installed at the Chashan Solar Radio Observatory (CSO), and solar radio observation tests were performed. The difference between the output of the system pointing at the cold sky versus the sun is approximately 3000 SFU, and the fluctuation of the radiation flux at each frequency point is small, basically within 50 SFU, when pointing at the sun on a clear and cloudless day. The output of the interferometer fluctuates by approximately 190 SFU when the observation is performed in weather with dense and thick clouds, while the radiation flux received by a single antenna at the same moment fluctuates by approximately 900 SFU. Therefore, the interferometer effectively mitigates the influence of the atmosphere, reduces the fluctuation of the system during observation, and improves the system sensitivity.

We still need to make improvements in the future. As shown in Figures 11 and 13, the amplitude-phase consistency of the system needs further calibration. Absolute calibration of the solar radiation flux is also needed. There are two distinct methods: blackbody calibration and signal source calibration. The former involves the placement of two blackbodies with identical characteristics on two separate antennas, allowing for precise and absolute calibration. The latter method entails the utilization of a signal source placed at a considerable distance from the system, whereby the system is pointed toward it for accurate calibration. In addition, by comparing the amplitude response of the output of the two-element interferometer of the tranquil Sun in Figure 12 with the varying input powers of the two-element interferometer in Figure 13, it is possible to determine the input power level of the two-element interferometer when receiving radiation signals from the tranquil Sun. The output amplitude of the system while observing the tranquil Sun is found to be 26, and based on Figure 13, the input power corresponding to an output amplitude of 26 is determined to be -78 dBm. However, it should be noted that this input power level is not situated within the completely linear range of the analog front end but rather within the small signal nonlinear range of the analog front end. To solve this problem, there are currently two measures: the first is to increase the input power of the analog front end by increasing the antenna diameter, and the second is to model and compensate for the nonlinear effect of the analog front end. The next stage will be to perform calibration work in these aspects.

\begin{acknowledgments}
This research is supported by the grants of National Natural Science Foundation of China (42127804) and the Qilu Young Researcher Project of Shandong University.
\end{acknowledgments}


\bibliography{sample631}
\bibliographystyle{aasjournal}


\end{CJK*}
\end{document}